\documentclass[twocolumn, trackchanges]{aastex6}

\shorttitle{Cooling timescale of torus in dying AGN}
\shortauthors{Ichikawa \& Tazaki}

\usepackage{amsmath}
\usepackage{color}
\usepackage{url}
\usepackage{ulem}
\usepackage{multirow}
\usepackage{longtable}

\newcommand{\rt}[1]{\textcolor{red}{#1}}

\usepackage{graphicx}
\usepackage{epstopdf}

\begin{document}


\title{Cooling timescale of dust tori in dying active galactic nuclei}


\author{Kohei~Ichikawa\altaffilmark{1,2,3} and Ryo~Tazaki\altaffilmark{4}
}
\affil{
$^1$ Department of Astronomy, Columbia University, 550 West 120th Street, New York, NY 10027, USA\\
$^2$ Department of Physics and Astronomy, University of Texas at San Antonio, One UTSA Circle, San Antonio, TX 78249, USA\\
$^3$ National Astronomical Observatory of Japan, 2-21-1 Osawa, Mitaka, Tokyo 181-8588, Japan\\
$^4$ Astronomical Institute, Tohoku University, 6-3 Aramaki, Aoba-ku, Sendai 980-8578, Japan
}

\email{k.ichikawa@astro.columbia.edu}



\begin{abstract}
We estimate the dust torus cooling timescale once
the active galactic nucleus (AGN) is quenched.
In a clumpy torus system, once the incoming photons are suppressed,
 the cooling timescale of one clump from $T_{\rm dust}=1000$~K to several 10~K 
is less than 10 years, indicating that the dust torus cooling time is
mainly governed by the light crossing time of the torus from
the central engine.
After considering the light crossing time of the torus, 
the AGN torus emission at 12~$\mu$m becomes over two orders
of magnitude fainter within $100$ years after the
quenching.
We also propose that those ``dying'' AGN
could be found using the AGN indicators with 
different physical scale $R$ such as 
12~$\mu$m band luminosity tracing AGN torus ($R \sim 10$~pc)
and the optical [OIII]$\lambda5007$ emission line narrow line regions
($R=10^{2-4}$~pc).
\end{abstract}





\section{INTRODUCTION}
Dust is the cornerstone of the unified view of active galactic nuclei (AGN).
The unified model of AGN \citep[e.g.,][]{ant85} proposes that all AGN are 
essentially the same;
all types of AGN have accretion disks, broad/narrow emission line regions, 
and those central engines are surrounded by optically and geometrically
thick dust ``tori'' \citep{kro86}.

Since the torus absorbs optical and ultraviolet photons from the 
accretion disk easily, the torus is heated and finally re-emits in
mid-infrared band.
X-ray emission also arises as inverse Compton scattering where
the source photons could originate from the accretion disk.
The strong luminosity correlations of AGN between hard X-ray 
and mid-infrared emission observationally support that
mid-infrared band is a good indicator of AGN torus emission
\citep[e.g.,][]{gan09, lev09, asm11, asm15, ich12, ich16b, mat15, gar16b}.

Recent observations, however, reported interesting populations of AGN. 
They show the AGN signatures in the larger physical scale with 
$>10^{2}-10^{5}$~pc scale \citep[e.g., narrow line regions; NLRs and/or
 radio jets; ][]{ben02,sch95},
but lack the AGN signatures in the smaller physical scales with
$<10$~pc (e.g., lack of X-ray emission, the emission from dust tori, and the radio cores).
This population is thought to be in the transient stage where 
they were active in the past, but now the central engine seems quiescent.
They are called fading AGN or dying AGN \citep[e.g.,][]{sch10, sch13, ich16a, sch16, men16, kee17}.

The large-scale AGN signature is also a good tool to 
constrain the AGN quenching time.
IC~2479 is one of the first fading AGN discovered through 
the galaxy zoo project \citep{lin08}.
\cite{lin09} first mentioned that high ionization lines including
[OIII]$\lambda$5007 are bright in the [OIII] blob,
while the [OIII] emission power around the AGN core is orders
of magnitude weaker, suggesting that AGN is fading.
\cite{sch10} confirmed this hypothesis through the X-ray 
observations with \textit{XMM-Newton} and \textit{Suzaku}.
Considering the distance from the central engine to the [OIII] 
blobs and input power, they estimated that the central 
engine of IC~2479 faded over two orders of magnitude 
within $10^4$ yrs.
\cite{ich16a} used a jet lobe size for estimating the upper
limit of the quenching time of the dying AGN.
Assuming a jet angle to the line of sight of $90^{\circ}$ and a typical expansion, the kinematic age of the radio jets is estimated to be $6\times10^{4}$~yr.
Therefore, the current understanding is that AGN reduce their luminosity over 
two orders of magnitude within $10^{4}$ yr, or even faster.

On the other hand, the size of the AGN dust torus is well suited
to our human timescale. 
Recent mid-infrared high spatial resolution
($\sim 0.3-0.7$ arcsec) observations have constrained 
the torus size of $<10$~pc \citep[e.g.,][]{pac05,mas06,
rad08, ram09, alo11,ich15}. 
Further, current mid-infrared interferometry observations have revealed
that a nearby AGN torus has a size of several pc 
\citep[e.g.,][]{jaf04,rab09,hon12,hon13,bur13,tri14,lop16}.
Detecting the decline of the torus emission compared to large-scale
feature such as NLR affords us to find a fresh quenching AGN within $\sim30$~yrs. 
To achieve this goal, it is crucial to estimate the torus cooling time 
quantitatively once the photon flux from the central engine becomes negligible.
In this paper, for the first step, we report an estimation of the cooling time of
AGN torus once the central engine is shut off in a simple assumption.

\section{Model}
Since the AGN torus is thought to be composed of a number of 
individual clumps,  the cooling timescale of the dust torus
is characterized by that of individual clumps unless the radiative 
interactions between  clumps are neglected. 
 As we discuss in Section \ref{sec:t_dep}, the radiative 
 interaction is expected to be less important in the cooling timescale. 
 In this Section, first, we describe the cooling timescale of 
 an individual clump and show that the clump cools down 
 with $\sim10$ year once the clump heating photons are lost. 
 Second, we calculate the spectral energy distribution (SED) 
of the AGN torus and estimate the attenuation of the flux in 
the mid-infrared wavelength as time goes by.
 
\subsection{Physical Properties of the Torus Clump}
We describe physical properties of the clump in the torus.
The idea of the torus clump is first proposed by \cite{kro88},
and then the model is sophisticated by many authors
\citep[e.g., ][]{bec04, hon07, nam14}.
The basic idea of the torus clump here is mainly compiled in \cite{vol04}.

The clump should be gravitationally bounded, and therefore the mass of the clump $M_c$ is thought to be larger than the Jeans mass $M_J$, 
\begin{equation}
M_{\rm c} \geq M_{J}\equiv\frac{\pi^{5/2}}{6}\frac{c_s^3}{G^{3/2}\rho_0^{1/2}} \label{eq:MJ}
\end{equation}
where $G$ is the gravitational constant, $c_s$ is the speed of sound, and $\rho_0$ is the gas density of clump.
In addition, the clump radius $R_{\rm c}$ should be smaller than the tidal radius $R_{\rm t}$, 
\begin{equation}
R_{\rm c}\leq R_{\rm t}\equiv\left(\frac{M_c}{3M_{\rm BH}}\right)^{\frac{1}{3}}r \label{eq:Rc}
\end{equation}
where $r$ is a distance from the black hole to the clump and $M_{\rm BH}$ is the black hole mass in the central engine.
Using $\rho_0=3M_c/4\pi R_c^3$, Equation (\ref{eq:MJ}) can be reduced to
\begin{equation}
M_c=\frac{\pi^2c_s^2}{3G}R_c. \label{eq:M2}
\end{equation}
Substituting Equation (\ref{eq:M2}) into Equation (\ref{eq:Rc}), we obtain
\begin{eqnarray}
R_c&=&\frac{\pi c_s}{3\sqrt{GM_{\rm BH}}}r^{3/2} \label{eq:Rc2}\\
&=& 4.8 \times 10^{-3} \left(\frac{c_s}{3\ {\rm km\ s}^{-1}}\right)\left(\frac{r}{1\ {\rm pc}}\right)^{\frac{3}{2}}\left(\frac{M_{\rm BH}}{10^8\ M_{\odot}}\right)^{-\frac{1}{2}} {\rm pc} 
\label{eq:Rc2_1}\nonumber\\
\end{eqnarray}
It follows from Equations (\ref{eq:M2}) and (\ref{eq:Rc2}) that the mass of marginally stable clump is
\begin{eqnarray}
M_{\rm c}&=&\frac{\pi^{3}}{9}\frac{c_s^3}{G^{3/2}M_{\rm BH}^{1/2}}\\
&=&33 \left(\frac{c_s}{3\ {\rm km\ s}^{-1}}\right)^3\left(\frac{r}{1\ {\rm pc}}\right)^{\frac{3}{2}}\left(\frac{M_{\rm BH}}{10^8\ M_{\odot}}\right)^{-\frac{1}{2}} M_{\odot}. \label{eq:Mc2_1}\nonumber\\
&&
\end{eqnarray}
The average hydrogen number density $n_{\rm H}$ of one clump
 is $n_{\rm H} = 3 M_{\rm c}/4 \pi R_{\rm c}^3 m_{\rm H}= 2.6 \times 10^9$~cm$^{-3}$, where $m_{\rm H}$ is the hydrogen mass. In this study, we assume the homogeneous spherical clump for simplicity. 
 
\subsection{Cooling Timescale of Individual Clump} \label{sec:clump}
Suppose the size distribution of dust grains obeys the power-law size distribution \citep{mat77, dra84},
\begin{equation}
n_i(a)da=A_in_{\rm H}a^{p}da \ \ (a_{\rm min}<a<a_{\rm max}) \label{eq:dn}
\end{equation}
where $n(a)$ is the distribution function of grain size, $a$ is a grain radius, $A$ is a normalization factor and $n_{\rm H}$ is the number density of H nuclei, and subscript $i$ denotes the silicate or the graphite.
Denote $E_{c,d}$ by the internal energy of a dust clump, then it can be described as follows: 
\begin{equation}
E_{c,d}(T_d)=M_{\rm c,g}f_d \frac{\sum_iA_i\int_0^{T_d} dT' C_i(T')}{\sum_iA_i\rho_i}, \label{eq:basic1}
\end{equation}
where $C(T_d)$ is the heat capacity per unit volume, $T$ is the temperature of a clump, $M_{\rm c,g}$ is the gas mass of dust clump, and $f_d$ is a dust-to-gas mass ratio. For the sake of simplicity, 
we assumed an individual clump has a uniform temperature.
It is noting that dust temperature at outer and inner regions within a clump may differ. However, this temperature difference does not significantly alter internal energy and cooling-curve of a clump. Hence, in this paper, we adopt single temperature approximation for an individual clump.

For the model of heat capacities, we adopt \citet{dra01}\footnote{There is a typographical error of Equation (10) of \citet{dra01} as pointed out in \citet{li02}. }, then 
\begin{eqnarray}
C_{\rm gra}&=&(N_C-2)k_B\left[f_2'\left(\frac{T}{863\ {\rm K}}\right)+2f_2'\left(\frac{T}{2504\ {\rm K}}\right)\right]\\
C_{\rm sil}&=&(N_A-2)k_B\left[2f_2'\left(\frac{T}{500\ {\rm K}}\right)+f_3'\left(\frac{T}{1500\ {\rm K}}\right)\right]
\end{eqnarray}
where 
\begin{equation}
f_n(x)=n\int_0^1 \frac{y^ndy}{\exp(y/x)-1},\ f_n'(x)=\frac{d}{dx}f_n(x).
\end{equation}
$N_C$ and $N_A$ is a number density of carbon and atoms for graphite and silicate, respectively.
Assuming $\rho_{\rm gra}=2.26$ g cm$^{-3}$ and $\rho_{\rm sil}=3.5$ g cm$^{-3}$ reduce to $N_C=1.5\times10^{23}$ cm$^{-3}$ and $N_A=8.5\times10^{22}$ cm$^{-3}$, respectively.

Next, we investigate how individual clump cools as time goes by.
If dust clump is optically thick, internal energy of the clump will be released through its photosphere, and then the cooling rate will be proportional to the surface area of the clump. Hence, we expect the cooling rate of optically thick clump approximately to be $4\pi R_c^2 \sigma_{\rm SB}T^4$, where $T$ is the temperature of clump and $\sigma_{\rm SB}$ is a Stefan-Boltzmann constant. On the other hand, if dust clump is optically thin, cooling rate should be the same as that of single dust grains.
Based on above consideration, we adopt the energy equation of individual clump for arbitrary optical depth as follows,
\begin{equation}
\frac{dE_{c,d}}{dt}= - 4\pi R_c^2 \langle Q_{\rm c}\rangle_T \sigma_{\rm SB} T^4 + \int_{V_{\mathrm{clump}}} n_g\Gamma_{\rm g-d}dV \label{eq:basic2}
\end{equation}
where $R_c$ is the radius of a clump, $n_g\Gamma_{\rm g-d}$ is the energy exchange between gas and dust, and $\langle Q_{\rm c} \rangle_T$
denotes the physical quantity $Q_{\rm c}$
to be averaged over the Planck function with 
temperature $T$. 
$Q_{\rm c}=[1-e^{-\tau_\nu}]$ is the emission efficiency of a 
clump, where $\tau_\nu$ is the optical depth for absorption of a clump defined by
\begin{equation}
\tau_\nu=\sum_i \int ds \int_{a_{\rm min}}^{a_{\rm max}}da n_i(a) m_{i}\kappa_{i,\nu}(a), \label{eq:tau}
\end{equation}
where $m$ is a mass of dust grain, and $\kappa_\nu(a)$ is the absorption opacity of dust grain.
From Equations (\ref{eq:tau}) and (\ref{eq:dn}), we obtain
\begin{equation}
\frac{\tau_\nu}{N_{\rm H}}=m_{\rm H}f_d\frac{\sum_{i} A_i\rho_i {\overline \kappa_{i, \nu}}}{\sum_i A_i\rho_i},
\end{equation}
 where $m_{\rm H}$ is the hydrogen mass, $f_d$ is the dust-to-gas mass ratio, and
\begin{equation}
{\overline \kappa_{i, \nu}} \equiv \frac{\int da a^{3+p} \kappa_{i, \nu}(a)}{\int da a^{3+p}}.
\end{equation}
The dust-to-gas mass ratio $f_d$ can be described as, for $p \neq -4$, 
\begin{equation}
f_{\rm d}=\frac{4\pi}{3m_{\rm H}}\frac{a_{\rm max}^{4+p}}{4+p}\left[1-\left(\frac{a_{\rm min}}{a_{\rm max}}\right)^{4+p}\right]\sum_{i} A_i\rho_{i}.
\end{equation}
The energy exchange between gas and dust can be written as
\begin{eqnarray}
n_g\Gamma_{\rm g-d}&=&\sum_i \int n_i(a)da4\pi a^2 \frac{n_gv_{\mathrm{th}}}{4}\alpha_T(2k_BT_g-2k_BT_d). \nonumber\\
\end{eqnarray}
Using Equation (\ref{eq:dn}), we find
\begin{eqnarray}
n_g\Gamma_{\rm g-d}&=&\frac{2\pi}{(3+p)}k_B\alpha_T(T_g-T_d)n_g^2v_{\rm th} \nonumber\\
&& a_{\rm min}^{3+p}\left[\left(\frac{a_{\rm max}}{a_{\rm min}}\right)^{3+p}-1\right]\sum_i A_i,
\end{eqnarray}
where $\alpha_T$ is the accommodation coefficient, $v_{\mathrm{th}}$ is the thermal velocity of gas, and $T_g$ is the gas temperature. We have averaged over the grain size distribution. In this paper, we assume $\alpha_T=0.15$ \citep[e.g.,][]{tie05}.
Since we have assumed uniform temperature and density structure in the clump, $n_g\Gamma_{\rm g-d}$ is constant within the cloud. Therefore, we obtain
\begin{eqnarray}
\frac{dT_d}{dt}&=&\beta\{-4\pi R_c^2 \langle Q_{\rm c}\rangle_{T_d} \sigma_{\rm SB} {T_d}^4 + n_g\Gamma_{\rm g-d}\cdot\frac{4\pi R_c^3}{3}\} \label{eq:dusttime} \\
\beta&=&\frac{1}{M_{c,g}f_{d}}\frac{\sum_i A_i\rho_i}{\sum_i A_iC_i(T_d)} \label{eq:dusttime2}.
\end{eqnarray}

The abundance ratio of silicates and graphites is set as
$A_{\rm sil}/A_{\rm gra}=1.12$ \citep{dra84}. 
This corresponds to 53\% of silicate grain, and the other
47\% for graphite in volume.
The upper and lower cut-off to the size distribution is
assumed to be $a_{\rm min}=0.005\ \mu$m and 
$a_{\rm max}=0.25\ \mu$m, and the slope adopted is
$p=-3.5$ \citep{mat77}. 
Since the dust-to-gas ratio $f_{\rm d}$ in AGN is still
under debate \citep[e.g.,][]{mai01}, we use the galactic 
ISM value of $f_{\rm d}=0.01$ \citep[e.g.,][]{tie05}
 throughout this paper. 
The dust grain is assumed to be spherical; thus, optical
properties can be calculated by using the Mie theory \citep[e.g.,][]{boh83}.
For the dielectric function of silicates and graphites,
we adopt \citet{dra84, lao93, wei01, dra03} and
\citet{dra84, ani12}, respectively.
Since graphite is a highly anisotropic material, 
optical properties depend on the direction of incident
${\bf E}$ fields with respect to the basal plane of graphite.
We assume randomly oriented graphite 
grains, and hence, we can use so-called ``$\frac{1}{3}-\frac{2}{3}$ approximation" \citep[e.g.,][]{dra88}; $\kappa_{{\rm gra}, \nu}=2\kappa_{{\rm gra}, \nu}({\bf E}\perp{\bf c})/3+\kappa_{{\rm gra}, \nu}({\bf E}\ ||\ {\bf c})/3$, 
where ${\bf E}$ is the incident electric field and ${\bf c}$ 
is a normal vector to the basal plane of graphite.
To estimate the free-electrons contribution to dielectric function for graphite with ${\bf E}\ ||\ {\bf c}$, we adopt the two-components free-electrons model introduced by \citet{ani12}. For free-electrons models of ${\bf E}\perp{\bf c}$, we still adopt the model of \citet{dra84}.

Since the gas-dust collision term depends on the gas temperature, the energy equation of gas should also be solved. 
Denote $E_{c,g}$ by the internal energy of the gas in the clump, then the energy equation for gas can be written as 
\begin{equation}
\frac{dE_{c,g}}{dt}=-\int_{V_{\mathrm{clump}}}\{n_g\Gamma_{\rm g-d}+n_g^2\Lambda_{\rm line}\}dV,
\end{equation}
where $n_g\Gamma_{\rm g-d}$ and $n_g^2\Lambda_{\rm line}$ represent the energy exchange between gas and dust and by line cooling, respectively. 
Note that $n_g\Gamma_{\rm g-d}$ has a positive sign when the gas temperature is higher than the dust temperature.
We have ignored the compression heating since free-fall timescale of the clump is longer than the cooling timescale of clump, e.g., $t_{\rm ff}=\sqrt{3\pi/32G\rho_g}=9.6\times10^{2}$ yr for the clump of $R_c\simeq5\times10^{-3}$ pc and $M_c\simeq 33\ M_\sun$.
The radiative cooling rate of through the transition from level $u$ to level $l$ of some species $x$ is written by \citep[e.g.,][]{tie85}
\begin{equation}
n_g^2\Lambda_x(\nu_{ul})=n_{u}A_{ul}h\nu_{ul}\beta(\tau_{ul})\{(S_x(\nu_{ul})-P(\nu_{ul}))/S_x(\nu_{ul}) \}
\end{equation}
where $n_u$ is a population density at level $u$, $\beta(\tau_{ul})$ is the espace probability, $S_x(\nu)$ is the source function, and $P(\nu)$ is the background radiation field. As a background radiation, we assume the ambient thermal radiation from dust grains, and then
\begin{equation}
P(\nu_{ul})=B(\nu_{ul},T_d)[1-\exp(-\tau_d)],
\end{equation}
where $\tau_d$ is calculated using Equation (\ref{eq:tau}).
Since we have assumed a homogeneous spherical clump, the 
escape probability averaged over line profile and over the cloud volume can be approximately given by \citep{dra11}
\begin{equation}
\beta(\tau_0) \simeq \frac{1}{1+0.5\tau_0}
\end{equation}
where $\tau_0$ is an optical depth at line center given by
\begin{equation}
\tau_0=\frac{g_u}{g_l}\frac{A_{ul}\lambda_{ul}^3}{4(2\pi)^{3/2}\sigma_V}n_lR_c\left(1-\frac{n_ug_l}{n_lg_u}\right),
\end{equation}
here we assumed that gas motion is assumed to be Maxwellian one dimensional distribution with the velocity dispersion of $\sigma_V=(k_BT_g/m_{\mathrm{H}})^{1/2}$. \\
Gas is assumed to be ideal, and hence, the equation of state becomes $e_{c,g}=n_gk_BT_g/(\gamma-1)$. As a result, we find
\begin{equation}
\frac{dT_g}{dt}=-\frac{(\gamma-1)}{n_g k_B}\{n_g\Gamma_{\rm g-d}+n_g^2\Lambda_{\rm line}\}, \label{eq:gastime}
\end{equation}
where $\gamma$ is a specific heat ratio and we adopt $\gamma=5/3$. 
In this paper, we assume the local thermal equilibrium (LTE) to obtain the level populations.
As molecular species, we adopt H$_2$, CO, and H$_2$O because these molecules are abundant in the AGN dust tori. Note that H$_2$O becomes abundant at relatively hot inner region of dust tori,
because H$_2$O molecules form via a neutral-neutral reaction whose energy barrier is $\simeq1000$~K, and the reaction is effective at $>300$~K \citep{har10}. 
We adopt the molecular fractional abundances with respect to $n_{\rm H}$ is $x_{\mathrm{H}_2}=0.5$,
 $x_{\mathrm{CO}}=3.0\times10^{-5}$, and $x_{\mathrm{H}_2\mathrm{O}}=1.4\times10^{-4}$ at 3 pc from the black hole \citep{har10}. We adopt the data of line parameters in the Leiden Atomic and Molecular Database LAMDA\footnote{\url{http://home.strw.leidenuniv.nl/~moldata/}} \citep{sch05} and H$_2$ line parameters in \citet{wag88, nom05}.
We use a part of the RATRAN code \footnote{\url{http://home.strw.leidenuniv.nl/~michiel/ratran/}} \citep{hog00} in order to read the molecular data for calculating the line cooling.

\begin{figure}[t]
\begin{center}
\includegraphics[height=6.6cm,keepaspectratio]{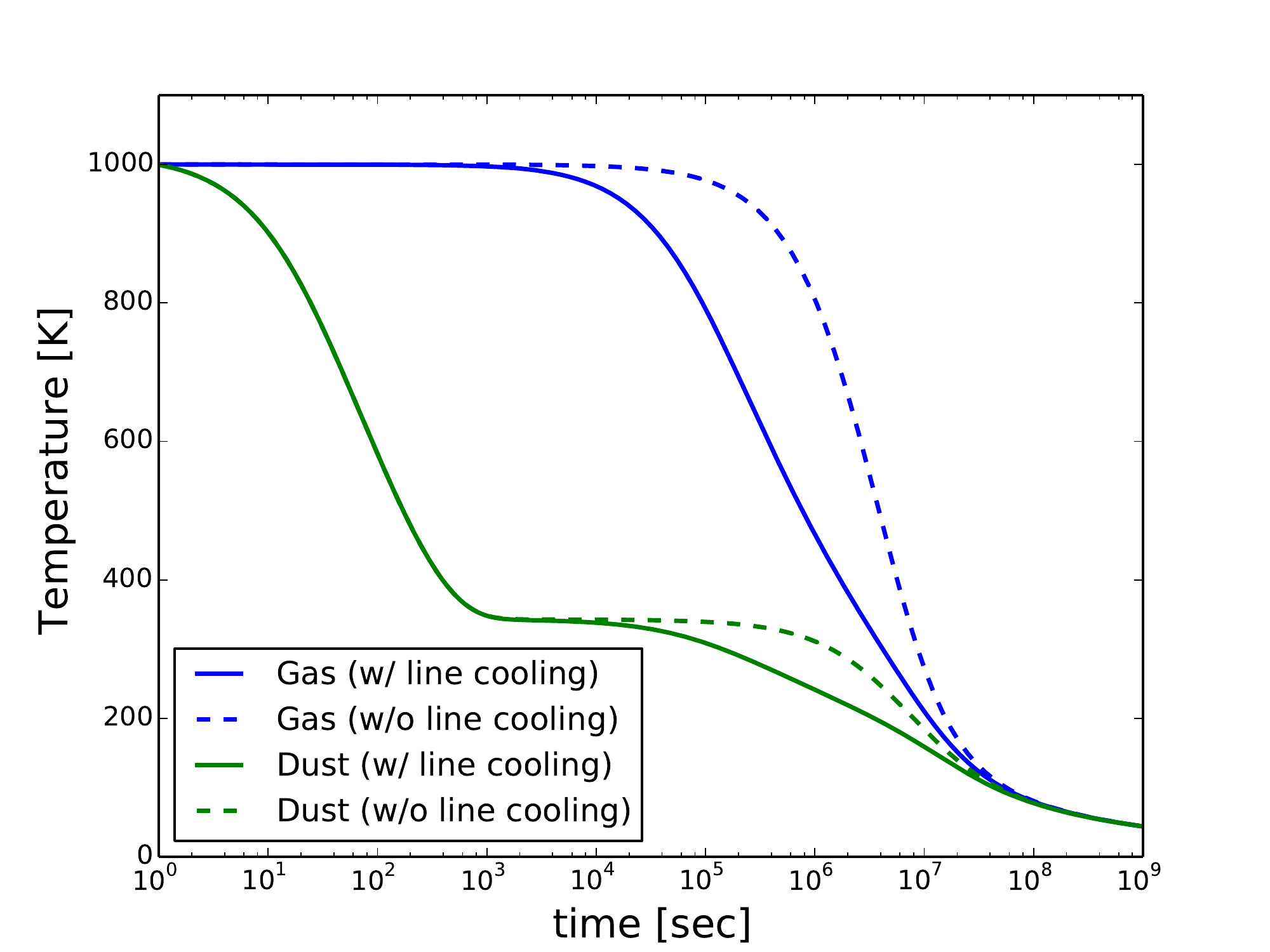}
\caption{Cooling function of individual clump. Solid and dashed lines represent models with and without molecular line cooling, respectively. Blue and green lines indicate the gas and dust temperature, respectively. Physical properties of the clump is derived assuming $r=1$ pc, 
and its initial gas and dust temperature are assumed to be 1000 K.}
\label{fig:cooling}
\end{center}
\end{figure}

In Figure \ref{fig:cooling}, we plot the change of dust and gas temperature 
of the individual clump. 
Figure \ref{fig:cooling} shows that the clump cools down to tens of 
Kelvin within $\sim10$ years. Dust temperature drops quickly until $t\sim 10^3$ s after the heating photon supply ends, and then it reaches a plateau phase where the radiative cooling and collisional heating of gas particles are balancing. Therefore, dust grains cannot cool down to tens of Kelvin unless the clump gas cools down. The gas temperature starts to decrease at $t\sim10^5$ s due to molecular line cooling that is mainly dominated by H$_2$O. Even in the absence of line cooling, gas clump will be cooled due to dust radiation via gas-dust collisions. As a result, dust temperature drops to tens of Kelvin within $\sim10$ years. It is worth noting that equilibrium dust temperature at the plateau phase depends on the gas density. This indicates that clumps at the outer region cool down faster than the inner clumps since they have lower gas density.
This result indicates that the cooling timescale of dust tori is slightly shorter than the light-crossing time with an order of 10~yr of the dust torus size. 
Therefore, we conclude that the dust cooling
timescale in the torus is mainly governed by the
light-crossing time.

\subsection{SED of Dust Tori}  \label{sec:sed}

To reproduce the torus SED, 
we assume smooth distribution of the dust density for simplicity.
The SED of AGN torus is obtained by using two-layers model
\citep{chi97, chi01}. 
In this model, the optically thick disk is divided into two regions,
the surface layer where the dust grains are directly irradiated by
the AGN, and interior layer where the dust grains are indirectly 
heated by the AGN, in other words, they are heated by the thermal
emission of directly irradiated dust grain at surface layer. 
The resultant SED can be obtained by superposing the flux from
the surface layer (Equation \ref{eq:FS}) and interior layer (Equation \ref{eq:FI}).

We assume the torus is axisymmetric, and the surface density 
of dust grains obeys the simple power law function, 
$\Sigma_{\rm d}=3.5\times10^{-2} (r/1\ {\rm pc})^{-1}$ g cm$^{-2}$. 
The inner and outer radii were observationally constrained 
and the AGN luminosity dependence was reported with
 $r_{\rm in}=0.4 (L_{\rm AGN}/10^{45}~{\rm erg~s{^{-1}}})^{0.5}$ pc
 \citep{bar87, kis11, kos14}, and 
$r_{\rm out}=10 (L_{\rm AGN}/10^{45}~{\rm erg~s{^{-1}}})^{0.21}$~pc \citep{kis11,gar16, ima16}, respectively. 
The opening angle of the torus is assumed to be 45 degrees, 
and then the grazing angle is $\alpha\approx r_{\rm in}/r$. 
The opacity of a mixture of silicate and graphite grains is 
obtained by averaging the absorption efficiency of them 
weighted for the volume fraction of 53\% and 47\% for
silicate and graphite, respectively.

The temperature of directly irradiated dust grains by the AGN is
\begin{equation}
T_{\rm ds}(r,a)=\left(\frac{L_{\rm AGN}}{16\pi r^2\sigma_{\rm SB}}\frac{\langle Q_{\rm abs} \rangle_{\rm AGN}}{\langle Q_{\rm abs} \rangle_{T_{\rm ds}}}\right)^{\frac{1}{4}} \label{eq:TDS}
\end{equation}
where $r$ is a distance from the AGN, and 
$L_{\rm AGN}=10^{45}$ erg s$^{-1}$ is the AGN luminosity.
$\langle Q_{\rm abs} \rangle_{\rm AGN}$ is an absorption efficiency
with respect to the incoming AGN photons, defined by
\begin{equation}
\langle Q_{\rm abs} \rangle_{\rm AGN}=\int Q_{\rm abs}f_\lambda d\lambda
\end{equation}
$f_{\lambda}=F_\lambda/F_{\rm AGN}$, and
$F_{\rm AGN}=\int_0^{\infty} F_\lambda d\lambda$ is the total flux. 
Using $F_{\rm AGN}=L_{\rm AGN}/4\pi r^2$, 
normalized spectrum of the AGN is assumed to be \citet{nen08a},
\begin{equation}
\lambda f_\lambda\propto\begin{cases}
\lambda^{1.2} & (\lambda \le \lambda_h) \\
\lambda^{0} & (\lambda_h \le \lambda \le \lambda_u) \\
\lambda^{-q} & (\lambda_u \le \lambda \le \lambda_{\rm RJ}) \\
\lambda^{-3} & (\lambda_{\rm RJ} \le \lambda) 
\end{cases}
\end{equation}
where $\lambda_h=0.01$~$\mu$m,  $\lambda_u=0.1$~$\mu$m, $\lambda_{\rm RJ}=1$~$\mu$m, and $q=0.5$.
The estimated dust temperature at the inner radius
is approximately $T_{\rm ds}\approx1500$ K which is consistent
with the sublimation temperature of the interior of a dust clump.
The temperature of interior grain is given by \citep{chi01},
\begin{equation}
T_{\rm di}(r)=\left\{\frac{L_{\rm AGN}}{8\pi r^2\sigma_{\rm SB}}\sin\alpha \frac{[1-e^{- \Sigma_d\langle\kappa\rangle_{T_{\rm ds}}}]}{[1-e^{-\Sigma_d\langle \kappa \rangle_{T_{\rm di}}}]}\right\}^{1/4} \label{eq:TDI}
\end{equation}
where $\kappa$ is a size distribution averaged grain opacity.
$\langle\kappa\rangle_{T_{\rm ds}}$ is evaluated at the temperature of most luminous grains in the surface.
We assume that the temperature of all grains at
midplane is thermally equilibrated.

\begin{figure}[t]
\begin{center}
\includegraphics[height=6.5cm,keepaspectratio]{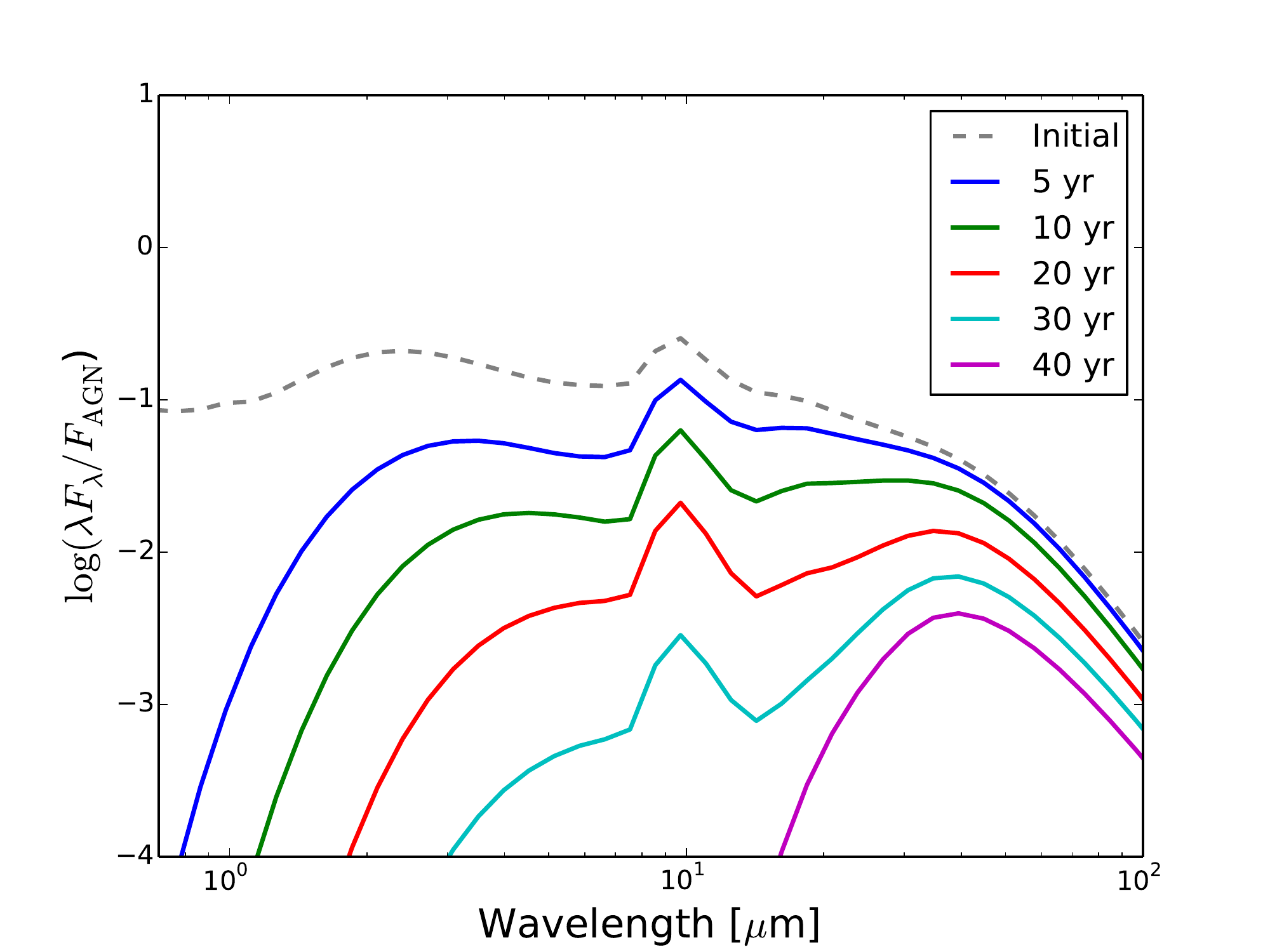}
\caption{Time evolution of dust torus SED 
around dying AGN whose initial luminosity
before the quenching is $10^{45}$~erg~s$^{-1}$.
The dashed line represents the SED under the radiative 
equilibrium temperature structure.}
\label{fig:sed}
\end{center}
\end{figure}

Once the radial distribution\rt{s} of grain temperature at the surface
and interior layer are obtained, the emission spectrum of AGN
torus can be calculated as below. 
The emission spectrum of the interior layer, $F_{\lambda}^{i}$, is 
\begin{equation}
4\pi d^2\lambda F_{\lambda}^{i}=8\pi^2\lambda\int_{r_{\rm in}}^{r_{\rm out}} B_\lambda(T_{\rm di})(1-e^{-\Sigma_{\rm d}\kappa}) rdr \label{eq:FS},
\end{equation}
where $d$ is the luminosity distance. 
The emission spectrum of the surface layer, $F_{\lambda}^{s}$, is 
\begin{equation}
4\pi d^2\lambda F_{\lambda}^{s}=8\pi^2\lambda\int_{r_{\rm in}}^{r_{\rm out}} (1+e^{-\Sigma_{\rm d}\kappa})S_\lambda(1-e^{-\tau_s}) rdr \label{eq:FI},
\end{equation}
where $S_\lambda$ and $\tau_s$ is the source function
and optical depth given by \citep[e.g.,][]{chi01},
\begin{eqnarray}
S_\lambda&=&\frac{2\int_{a_{\rm min}}^{a_{\rm max}}B_\lambda(T_{\rm ds})n(a)a^2 Q_{\rm abs}(a,\lambda)da}{\int_{a_{\rm min}}^{a_{\rm max}} n(a)a^2 Q_{\rm abs}(a,\lambda)da}\\
\tau_s&=&\frac{\int_{a_{\rm min}}^{a_{\rm max}}n(a)a^2 Q_{\rm abs}(a,\lambda)da}{\int_{a_{\rm min}}^{a_{\rm max}} n(a)a^2 \langle Q_{\rm abs} \rangle_{\rm AGN} da}\sin\alpha.
\end{eqnarray}
The dashed gray line in the Figure \ref{fig:sed} shows 
the SED for the equilibrium temperature structure of the dust torus.

\begin{figure}[t]
\begin{center}
\includegraphics[height=6.8cm,keepaspectratio]{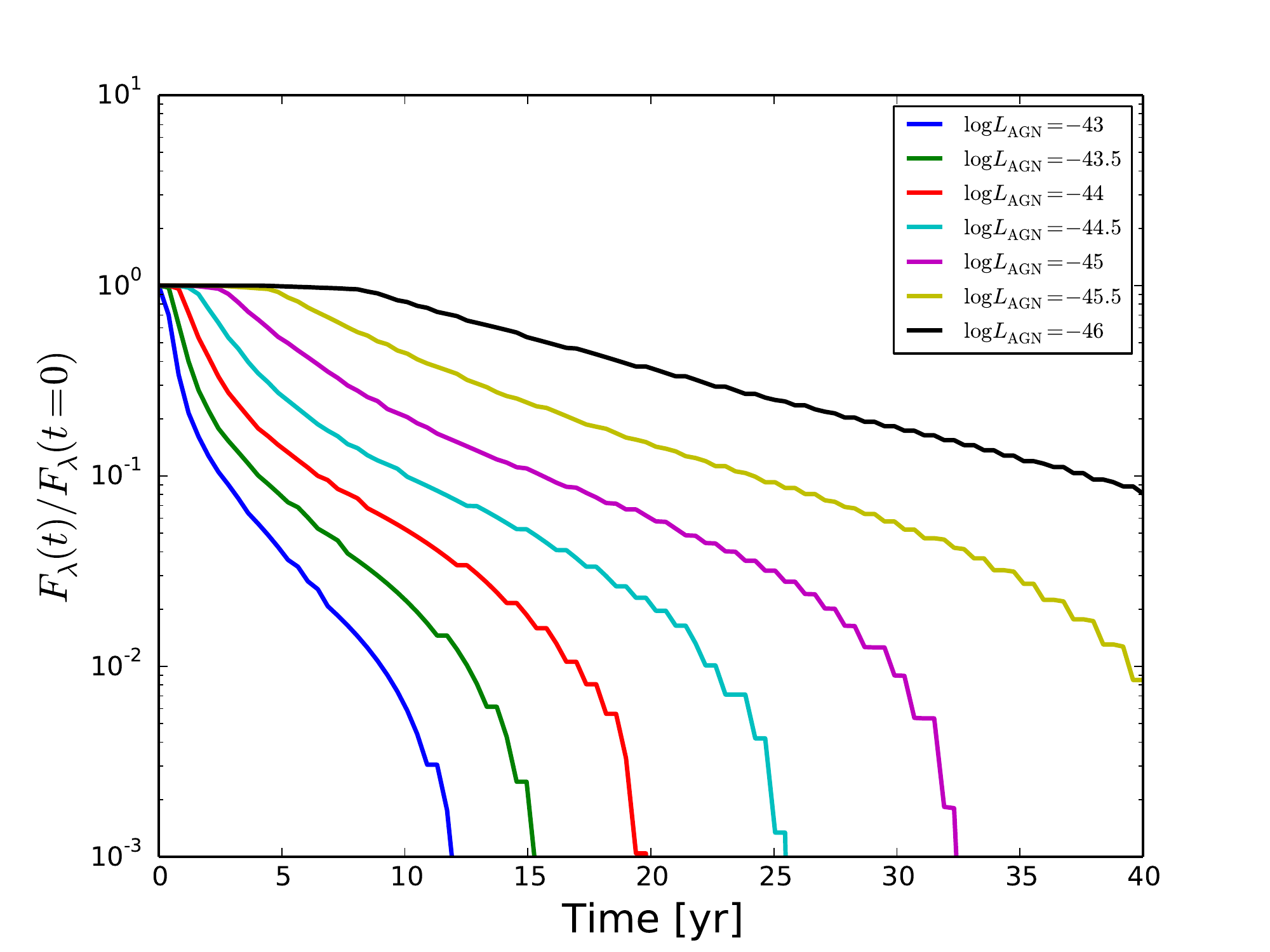}
\caption{The decline of AGN $12 \mu$m luminosity normalized by the initial ($t=0$) AGN $12 \mu$m luminosity as a function of time since the AGN is quenched. }
\label{fig:sed12um}
\end{center}
\end{figure}


\section{Discussion}

\subsection{Time Dependence of SED}\label{sec:t_dep}
The clump is optically thick in the shorter wavelength,
and optically thin in the longer wavelength than mid-infrared
emission domain.
The dust clump at midplane is heated by the near-infrared
emission of dust clump at the surface layer and reradiates
its energy in mid- to far-infrared wavelength.
 In far-infrared wavelength, since a clump is optically thin,
 most of photons emitted from the midplane escape without 
 experiencing significant absorption.
 At the surface optically thin layer, clumps are sparsely distributed, 
 and hence, the radiative interaction between clumps at the 
 surface layer does not frequently occur. 
Based on above considerations, 
we assume that the cooling timescale of the torus
is mainly governed by that of each clump.

We calculate the time evolution of the torus SED after the central AGN is quenched. 
We assume that the temperature of dust tori given by
Equations (\ref{eq:TDS}) and (\ref{eq:TDI}) will be cooled 
according the characteristic cooling-curve of individual 
clump defined by Equations (\ref{eq:dusttime}), (\ref{eq:dusttime2}), and (\ref{eq:gastime}). 
In addition, to take into account the light crossing time, the surface and midplane temperature at radius $r$ starts to decrease at $t=r/c$ and $t=2r/c$ after the AGN is quenched, respectively. The factor $2$ in the latter is inserted so that the light crossing time of vertical direction is taken into account as well as the radial direction. Since cooling timescale depends on the clump density, or the black hole mass, 
we derive the black hole mass by setting that
the initial AGN luminosity is 5 \% of the 
Eddington luminosity \citep[e.g.,][]{kel10}.

Figure \ref{fig:sed} shows the time dependence of the AGN torus SED with face-on view after the quenching of the central engine with the initial AGN luminosity
of $L_{\rm AGN}=10^{45}$~erg~s$^{-1}$.
Once the high-energy photons stop being emitted, 
individual clumps cool down rapidly, and the light crossing
time of the torus, $r_{\rm out}/c\approx 30$~yr, governs
the cooling timescale. 
In Figure \ref{fig:sed12um}, we plot the change of mid-infrared
12~$\mu$m emission flux as a function of time since the central engine is quenched.
The torus in higher luminosity AGN shows a longer cooling
time because of the larger torus outer radius as shown in Section~\ref{sec:sed}.

\begin{figure}[t]
\begin{center}
\includegraphics[height=6.8cm,keepaspectratio]{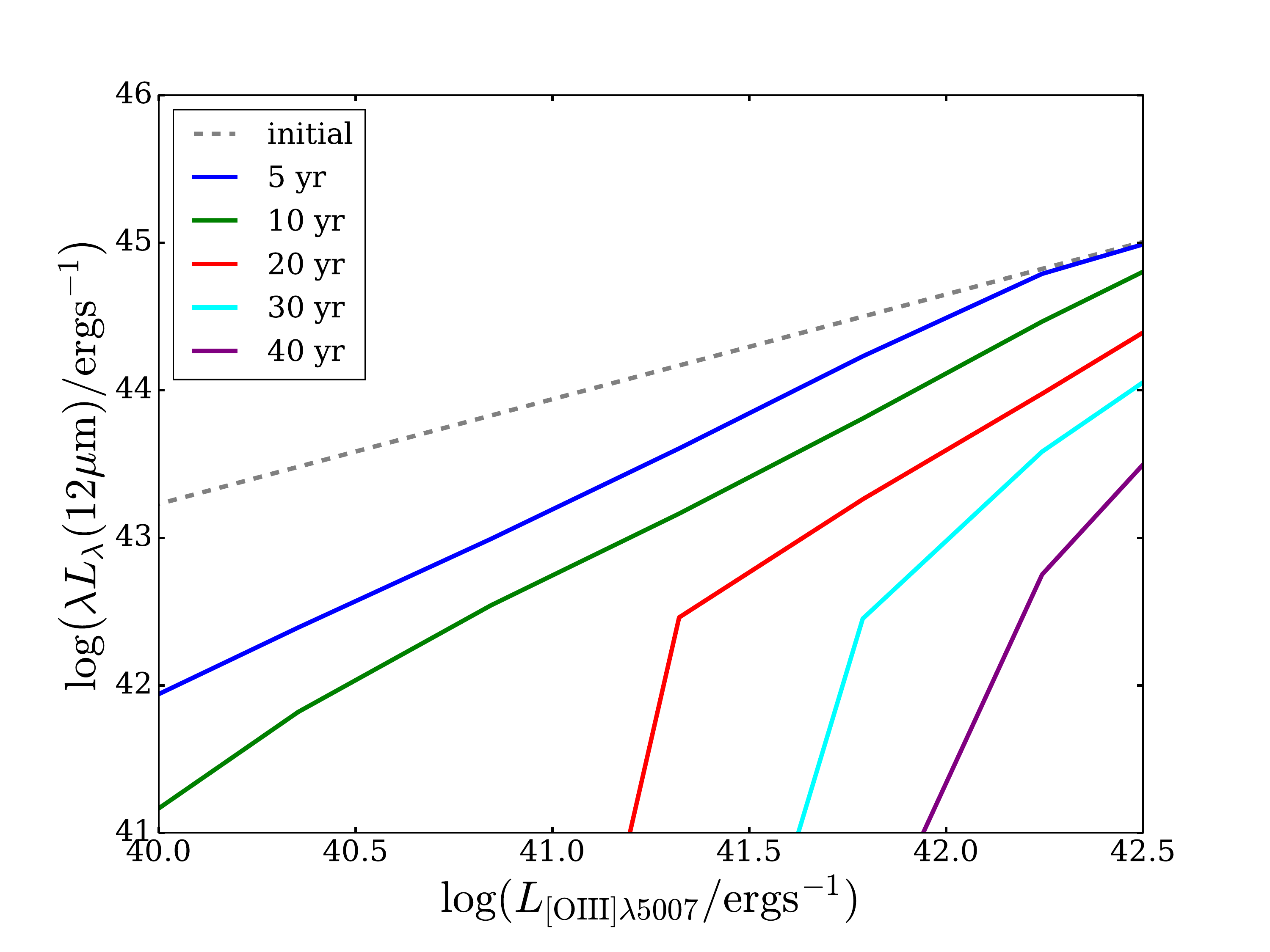}
\caption{Time evolution of the AGN luminosity relation between 12~$\mu$m and [OIII]$\lambda5007$. }
\label{fig:lmirvslo3}
\end{center}
\end{figure}


\subsection{Searching for Dying AGN Candidates}
To search for dying AGN candidates discussed above, 
a longer and more stable timescale AGN indicator than
the torus thermal emission is necessary to compare with the
 torus mid-infrared luminosity.
NLR is a promising tool because their size is generally
$10^{2-4}$~pc, and [OIII]$\lambda5007$ emission
line is one of the good
NLR indicators hosting a strong luminosity correlation with AGN
indicators \citep{ued15}.
A promising method is to cross-match optically selected type-2 AGN obtained 
by the Sloan Digital Sky Survey \citep[SDSS;][]{yor00} 
with \textsc{Allwise} catalogs \citep{wri10}.
SDSS type-2 AGN have prominent [OIII]$\lambda5007$ emission and \textsc{Allwise} will give us MIR 12 or 22~$\mu$m emission.
Based on the luminosity relations of AGN between [OIII]$\lambda5007$ and 12~$\mu$m luminosity \citep{tob14}, 
$\log (L_{\rm [OIII]}/{\rm erg~s}^{-1}) \sim 41$ is equivalent to $\log (L_{{\rm 12~}\mu{\rm m}}/{\rm erg~s}^{-1}) \sim 44$, which is as luminous as QSO as shown
with the gray dashed line in Figure~\ref{fig:lmirvslo3}.
Using the time evolution function of 12~$\mu$m luminosity in Figure~\ref{fig:sed12um},
we calculate the time evolution of the luminosity relation of [OIII]$\lambda5007$ and 12~$\mu$m luminosity in Figure~\ref{fig:lmirvslo3}. 
The AGN locating at the bottom right area in Figure~\ref{fig:lmirvslo3} would be a prominent candidate of those dying AGN.
This study could be achievable for type-2 AGN with the redshift range of $z<0.3$, 
where the optical line diagnostics of AGN can be applied \citep[e.g.,][]{kew06}.
Further studies using the method above will be discussed in a forthcoming paper.

The ongoing Subaru HSC \citep{miy12} SSP deep survey \citep{aih17a, aih17b} covering 28~deg$^{2}$ will also be a promising region for finding dying AGN using [OIII] emitters at $z \sim 0.6$ and $\sim 0.8$ obtained with a narrow band filter NB816, NB921, respectively. Again, if there are sources with extremely low ratio of $L_{12,22\mu{\rm m}}/L_{\rm [OIII]}$, this could be a prominent candidate of high-$z$ dying AGN which cannot be covered with the method of SDSS survey above. In this case, optical or near-infrared spectral follow-up is crucial to disentangle the [OIII] emitters into starburst galaxies and AGN \citep[e.g.,][]{kew06}.

\vspace{3mm}
\section{Conclusions}
The cooling time of the torus is mainly governed by the light
crossing time of the torus from the central engine.
The dust torus cools down by roughly 1 orders of magnitude within 10~yr once the propagation of
the photon from the central engine stops, and the
dust torus emission completely disappears
with $<100$~yr for most of the AGN luminosity range
as shown in Figure~\ref{fig:sed12um}.
Those weak dust-torus emissions or ``dying'' AGN could be found 
with the combination of the optical spectral or 
narrow band survey detecting the NLR indicator [OIII]$\lambda$5007
by cross-matching with \textsc{Allwise} 12 or 22~$\mu$m band.




\acknowledgments
We would like to acknowledge the anonymous
referee for useful suggestions that have helped
to clarify this paper.
We would like to thank Hideko Nomura for providing useful feedback 
to improve initial model set-up and gas line cooling.
We also thank Takuma Izumi, Toshihiro Kawaguchi, Yoshiki Matsuoka, Takeo Minezaki, and Chris Packham for fruitful discussions.
We thank Bruce T. Draine for providing the data of dielectric function of graphite.
This work was partly supported by the Grant-in-Aid for Scientific Research 40756293 (K.I.)
and by the Grant-in-Aid for JSPS fellow for young researchers (PD: K.I. and R.T.).

\end{document}